# Dark matter searches

*Laura Baudis*[1, *]

One of the major challenges of modern physics is to decipher the nature of dark matter. Astrophysical observations provide ample evidence for the existence of an invisible and dominant mass component in the observable universe, from the scales of galaxies up to the largest cosmological scales. The dark matter could be made of new, yet undiscovered elementary particles, with allowed masses and interaction strengths with normal matter spanning an enormous range. Axions, produced non-thermally in the early universe, and weakly interacting massive particles (WIMPs), which froze out of thermal equilibrium with a relic density matching the observations, represent two well-motivated, generic classes of dark matter candidates. Dark matter axions could be detected by exploiting their predicted coupling to two photons, where the highest sensitivity is reached by experiments using a microwave cavity permeated by a strong magnetic field. WIMPs could be directly observed via scatters off atomic nuclei in underground, ultra low-background detectors, or indirectly, via secondary radiation produced when they pair annihilate. They could also be generated at particle colliders such as the LHC, where associated particles produced in the same process are to be detected. After a brief motivation and an introduction to the phenomenology of particle dark matter detection, I will discuss the most promising experimental techniques to search for axions and WIMPs, addressing their current and future science reach, as well as their complementarity.

## 1 Introduction

After many decades of increasingly precise astrophysical observations, we have unequivocal evidence that the majority of the material that forms galaxies, clusters of galaxies and the largest observed structures in the cosmos is non-luminous, or dark. This conclusion rests upon accurate measurements of galactic rotation curves, measurements of orbital velocities of individual galaxies in clusters, cluster mass determinations via gravitational lensing, precise measurements of the cosmic microwave background acoustic fluctuations and of the abundance of light elements, and upon the mapping of large scale structures. In addition, cosmological simulations based on the ΛCDM model successfully predict the observed large-scale structures in the universe. In this model, which today provides the only paradigm that can explain all observations, our universe is spatially flat and composed of ∼5% atoms, ∼27% dark matter and ∼68% dark energy [1].

The first quantitative case for a dark matter dominance of the Coma galaxy cluster was made as early as 1933 by the Swiss astronomer Fritz Zwicky [2]. Since then, our understanding of the amount and distribution of dark matter on various scales, including the Milky Way, deepened, but we still lack the answer to the most basic question: what is the dark matter made of? One intriguing answer is that it is made of new particles, yet to be discovered. Instantly, more questions arise: what are the properties of these particles, such as their mass, interaction cross section, spin and other quantum numbers? Is it one particle species, or many? Are these particles absolutely stable, or very long-lived? I will discuss searches for two particular classes of dark matter particle candidates, motivated independently by problems within the Standard Model of particle physics. These are QCD axions with masses in the range ∼1 $\mu$eV - 10 meV and weakly interacting massive particles (WIMPs) with masses in the ∼0.3 GeV-100 TeV range.

## 2 Axions

The axion is one of the most promising hypothetical particles proposed to solve the dark matter puzzle [3]. Originally, axions were introduced by Peccei and Quinn (PQ) as a solution to the strong-CP problem in QCD [4]. They postulated a global U(1) symmetry that is spontaneously

---

[*] Corresponding author   E-mail: laura.baudis@uzh.ch
[1] Physik Institut, University of Zurich, Switzerland

broken below an energy scale $f_a$. While the original PQ axion with $f_a$ around the electroweak symmetry-breaking scale was soon excluded, so-called invisible axion models, with $f_a$ much larger than the weak scale are still viable [5]. The axion mass and decay constant $f_a$ are related to those of the pion, $m_a f_a \approx m_\pi f_\pi$, where $m_\pi$=135 MeV and $f_\pi \approx$ 92 MeV. More precisely, the relation is [6]:

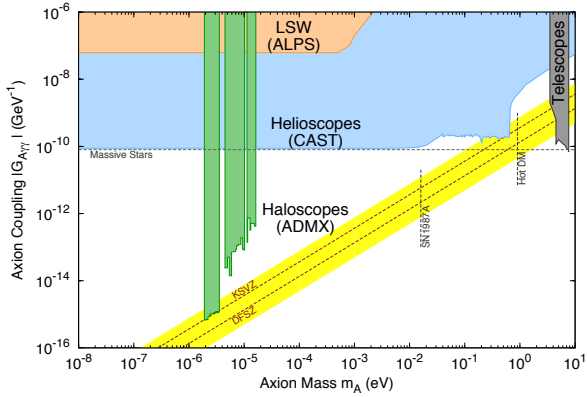

**Figure 1** Axion coupling versus axion mass parameter space, where generic axion models are shown as a yellow band (see equation 1). The current sensitivity of various axion searches, including the ADMX experiment [7], is shown. Figure from [6].

$$m_a = \frac{\sqrt{m_u/m_d}}{1+m_u/m_d} \frac{f_\pi m_\pi}{f_a} \approx 6 \cdot 10^{-6} \text{eV} \frac{10^{12}\,\text{GeV}}{f_a}. \quad (1)$$

In more general axion-like particle (ALP) models, the two parameters, namely the ALP mass and its decay constant are independent of one another, opening up a much larger parameter space for searches, as shown in Figure 1.

Cold, QCD axions are produced during the QCD phase transition, but the exact mass of the axion is not known. Generic axion couplings are proportional to $f_\pi/f_a$, however large model-dependencies and uncertainties exist [6]. For dark matter searches, the axion coupling to two photons, dictated by the following Lagrangian, is of particular interest:

$$\mathcal{L}_{a\gamma\gamma} = -g_\gamma \frac{\alpha}{\pi f_a} a(x) \mathbf{E}(x) \cdot \mathbf{B}(x) = -g_{a\gamma\gamma} a(x) \mathbf{E}(x) \cdot \mathbf{B}(x), \quad (2)$$

where $a(x)$ is the scalar axion field, $\mathbf{E}(x)$ and $\mathbf{B}(x)$ are the electric and magnetic field of the two propagating photons, $\alpha$ is the fine structure constant and $g_\gamma$ is a model-dependent coefficient of order one ($g_\gamma$ =0.36 and $g_\gamma$=-0.97 in the generic DFSZ [8] and KSVZ [9] axion models, respectively).

The coupling of the axion to electromagnetism is inferred to be extremely weak, and the two-photon decay width is:

$$\Gamma_{a \to \gamma\gamma} = \frac{g_{a\gamma\gamma}^2 m_a^3}{64\pi} = 1.1 \times 10^{-24} s^{-1} \left(\frac{m_a}{\text{eV}}\right)^5. \quad (3)$$

Thus the spontaneous decay life-time of an axion to two real photons is vastly greater than the age of our universe. We can also see from the above relation that the axion decays faster than the age of the universe if $m_a \gtrsim$ 20 eV.

Notwithstanding, axions can be detected through axion-photon conversion in external electric or magnetic field [10], and a variety of laboratory searches exploit this approach (see [6] for a recent review). Constraints from such searches and from astrophysical observations restrict the axion mass range to $\sim$1 $\mu$eV - 10 meV [11]. The upper bound is rather robust, and comes from the fact that axions can be produced in astrophysical bodies and escape, leading to new sources of energy loss. The lower bound comes from the requirement that the relic density of the axion is smaller or equal to the observed dark matter density, $\Omega_{dm}$. However, the magnitude of this bound depends on whether inflation occurred before or after the PQ phase transition and on the thermal history of our universe, see e.g. [12, 13]. Thus, in contrast to WIMPs, which I will discuss in the next section, axions do not naturally have the correct relic density to provide the dark matter, as their mass $m_a$ (and thus coupling) can vary over a wide range.

Dark matter axions can in principle be detected via the inverse Primakoff effect, where the axion decay is accelerated through a static, external magnetic field. In this field, one photon is replaced by a virtual photon, and the other maintains the energy of the axion, namely its rest mass plus its kinetic energy:

$$E \simeq m_a c^2 + \frac{1}{2} m_a c^2 \beta^2. \quad (4)$$

The ADMX experiment [7] exploits the axion detection scheme proposed by Sikivie [10], based on the Primakoff effect: in a microwave cavity permeated by a strong magnetic field, the axion-photon conversion is enhanced when the resonant frequency $f$ of the cavity equals the axion rest mass:

$$f \simeq \frac{m_a c^2}{h} \quad (5)$$

where h is Planck's constant. Assuming an axion mass of 5 $\mu$eV and an expected velocity dispersion of $\Delta\beta \sim 10^{-3}$ in our galaxy, the spread in the axion energy is expected to be $\sim 10^{-6}$ or $\sim 1.2$ kHz. The resonant cavity is tuneable, and the axion decay signal is to be detected by observing the proper modes at a given frequency. The expected axion-to-photon conversion power is:

$$P = g_{a\gamma\gamma}^2 \frac{\rho_a}{m_a} B_0^2 V C \min(Q_L, Q_a)$$
$$= 4 \times 10^{-26} \text{W} \left(\frac{g_\gamma}{0.97}\right)^2 \frac{\rho_a}{0.5 \times 10^{-24} \text{g cm}^{-3}} \frac{m_a}{2\pi(\text{GHz})} \quad (6)$$
$$\times \left(\frac{B_0}{8.5 \text{T}}\right)^2 \frac{V}{0.22 \text{m}^3} C \min(Q_L, Q_a),$$

where $\rho_a$ is the local axion density, $B_0$ is the strength of the static magnetic field, V is the cavity volume, C is a mode-dependent cavity form factor, and $\min(Q_L, Q_a)$ is the smaller of either the cavity or axion quality factors. The axion signal quality factor is $Q_a = 10^6$, the ratio of their energy to the energy spread. The signal power is thus expected to be exceedingly weak, and the cavity and amplifiers are cooled to very low temperatures to minimize thermal noise.

While ADMX had already started to probe part of the QCD axion parameter space, as shown in Figure 1, the new ADMX experiment is currently being assembled at the University of Washington. It will use a tuneable cavity with a higher quality factor, and a lower intrinsic noise, due to a dilution refrigerator and quantum-limited SQUID amplifiers. Starting in late 2015, ADMX will test a sizeable fraction of the predicted parameter space for the QCD axion as a dark matter candidate. In addition, an R&D program for a next-generation experiment (ADMX-HF), that can probe the theoretically allowed higher mass region (10-100 GHz) is ongoing. A dark matter axion experiment based on a large-volume, high magnetic field and a high quality microwave cavity is in planning at the Korean IBS Center for Axion and Precision Physics Research [14].

## 3 WIMPs: overview

Weakly interactive massive particles (WIMPs), which would have been in thermal equilibrium with quarks and leptons in the hot early universe, and decoupled when they were non-relativistic, represent a generic class of cold dark matter candidates [15]. Their relic density can account for the dark matter density if the annihilation cross section $\sigma_{ann}$ is around the weak scale:

$$\Omega h^2 \simeq 3 \times 10^{-27} \text{cm}^3 \text{s}^{-1} \frac{1}{\langle \sigma_{ann} v \rangle} \quad (7)$$

where $v$ is the relative velocity of the annihilating WIMPs, and the angle brackets denote an average over the WIMP thermal distribution. $\langle \sigma_{ann} v \rangle = 3 \times 10^{-26} \text{cm}^2 \text{s}^{-1}$ gives the correct relic density, and is often considered as the benchmark annihilation cross section value. Concrete examples for WIMPs are the lightest superpartner in supersymmetry with R-parity conservation [16], and the lightest Kaluza-Klein particle, for instance the first excitation of the hypercharge gauge boson, in theories with universal extra dimensions [17, 18].

Perhaps the most intriguing aspect of the WIMP hypothesis is the fact that it is testable by experiment. WIMPs with masses around the electroweak scale are within reach of high-energy colliders and of direct and indirect dark matter searches [19]. Direct detection experiments search for a scattering signal between the WIMP and an atomic nucleus, while indirect detection experiments aim to detect annihilation products of dark matter particles, such as neutrinos, gamma rays, antiprotons and positrons above the astrophysical background. Colliders look for dark matter production in high energy particle collisions. At the LHC, the signature is missing energy, as the WIMP leaves the detectors unobserved, accompanied by a jet, a photon, a Z etc, that are required for tagging an event. These approaches are complementary to one another, and to astrophysical probes that can test non-gravitational interactions of dark matter [20]. Although no conclusive evidence for WIMPs in any of these detection channels exists, it is likely that we will be faced with detections in multiple approaches within this decade.

## 4 Direct WIMP detection

The local density of dark matter is around $7 \times 10^{-25}$ g/cm$^3$ (0.4 GeV/cm$^3$) [21, 22], implying a WIMP flux onto the Earth of about $10^5$ cm$^{-2}$s$^{-1}$, for a particle mass of $\sim$100 GeV. This allows us to in principle detect WIMPs as they pass through and scatter elastically off nuclei in a terrestrial detector [23]. Considering a WIMP velocity dispersion of $\sigma \approx 270$ km s$^{-1}$ and an escape velocity of v$_{esc} \approx 544$ km s$^{-1}$ [24, 25] in the galactic rest frame, a dark matter particle with a mass in the GeV–TeV range has a mean momentum of a few tens of MeV and an energy below 50 keV is transferred to the nucleus during the collision. Expected event rates range from about one event to less than $10^{-3}$ events per kilogram detector material and year. Thus, to observe a WIMP-induced nuclear recoil spectrum, a low energy threshold, an ultra-low background noise and a relatively large target mass are essential. Distinct experimental signatures from a particle populating our galactic halo come from the Earth's motion through the galaxy. It induces both a seasonal variation of the total event rate [26, 27] and a forward-backward asymmetry in a directional signal [28, 29].

A variety of techniques are employed to search for the tiny signal that arises when the energy of the scattered nucleus is transformed into ionisation, scintillation light or phonons. Among those with highest sensitivities are cryogenic experiments operated at sub-Kelvin temperatures, detectors based on liquified argon or xenon and superheated liquid detectors. In addition, a strong R&D program for detectors capable of measuring the direction of the recoiling nucleus is ongoing [30]. The main background sources for these experiments are radio-impurities in the detector construction materials, neutrons from $(\alpha, n)$ and fission reactions, cosmic rays and their secondaries, activation of detector materials during exposure at the Earth's surface, as well as sources intrinsic to the target materials. The ultimate backgrounds might come from neutrino-induced nuclear recoils from coherent neutrino-nucleus scatters. The coherent scattering rate from $^8$B solar neutrinos will provide an irreducible background for low-mass WIMPs, limiting the cross section sensitivity to $\sim 4 \times 10^{-45}$ cm$^2$ for WIMPs of 6 GeV/$c^2$ mass. Atmospheric and diffuse supernova neutrinos will dominate the rate for WIMP masses above $\sim 10$ GeV/$c^2$ and cross section below $10^{-49}$ cm$^2$ [31–35]. This background can be in principle overcome by experiments with directional sensitivity, even for those detectors that would only measure 1-d or 2-d projections of the recoil tracks, see e.g. Ref. [36].

Current experimental upper limits and projected sensitivities for the spin-independent WIMP-nucleon interactions as a function of the WIMP mass are summarised in Figure 2. In spite of observed anomalies in a handful of experiments, that were interpreted as due to WIMPs, we have no convincing evidence of a direct detection signal induced by galactic dark matter. The low mass region is accessible by detectors with very low energy threshold and/or lighter target nuclei, while the higher mass region is probed by experiments with very low background rates and high target masses. Below WIMP masses of 6 GeV, the tightest constraints are derived from DAMIC [37], CRESST [38], EDELWEISS [39] and SuperCDMS [40].

The most constraining upper limits for WIMP masses above 6 GeV come from the LUX experiment [41] using a liquid xenon time-projection chamber. The reached 90% upper C.L. cross section for spin-independent WIMP-nucleon couplings has a minimum of $7.6 \times 10^{-46}$ cm$^2$ at a WIMP mass of 33 GeV/$c^2$, thus confirming and improving upon the previous XENON100 results [42]. Several new, ton-scale experiments are either in commissioning (ArDM [43], DEAP-3600 [44]), in construction (XENON1T [45]) or in advanced planning phase (LZ [46], XENONnT [47], XMASS-5t [48]) in underground laboratories around the world, with aimed sensitivities which are a factor of 40-100 higher than current ones. Finally, so-called ultimate WIMP detectors such as DARWIN [35, 49], operating multi-tons of target material, would have the capability of probing the entire experimentally accessible parameter space, until the neutrino-induced background will eventually start to become visible (see Figure 2).

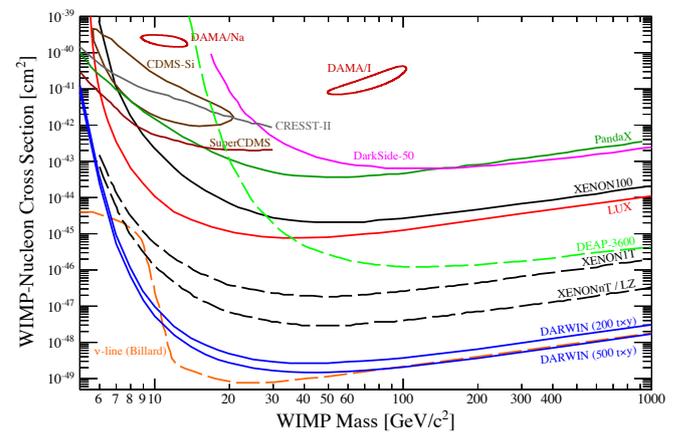

**Figure 2** Spin-independent WIMP-nucleon scattering results: Existing upper limits from the CRESST-II [38], SuperCDMS [40], PandaX [50], DarkSide-50 [51], XENON100 [42], and LUX [41] experiments, along with projections for DEAP3600 [44], XENON1T [45], XENONnT [47], LZ [46], and DARWIN [49] are shown. DARWIN is designed to probe the entire parameter region for WIMP masses above $\sim 6$ GeV/$c^2$, until the neutrino background ($\nu$-line) will start to dominate the recoil spectrum. Figure adapted from [52].

Recent estimates of theoretically allowed regions for the WIMP-nucleon cross section in extensions of the Standard Model containing dark matter candidates range from a few $\times 10^{-45}$ to about $1 \times 10^{-52}$ cm$^2$, depending on the theoretical framework, its underlying assumptions and the number of free parameters. As a concrete example, in the constrained minimal supersymmetric standard model (CMSSM), the favourite dark matter candidate is an almost pure higgsino neutralino with a mass around 1 TeV [53] and a spin-independent cross section on the nucleon between $\sim 10^{-46}$ cm$^2$ and a few $\times 10^{-45}$ cm$^2$. This high-mass region, which takes into account the Higgs mass of $\sim 125$ GeV, lower limits from direct SUSY searches at the LHC, and a relic WIMP abundance of $\Omega_{DM}h^2 = 0.12$, is well within the reach of the DEAP-3600, SuperCDMS, and XENON1T experiments. More generally, the relatively large Higgs mass pushes the WIMP mass and cross section to higher and lower values, respectively, due to the higher value of the SUSY breaking scale, as shown in a number of studies [54–56].

In more general, phenomenological MSSM models, with 15-19 free parameters, additional regions in the parameter space open up, where the neutralino could be bino-, higgsino-, or wino-like, with cross sections extending down to ∼ $10^{-51} – 10^{-52}$ cm$^2$ or even lower [57–59]. While part of this parameter space is within the reach of XENON1T, or the future XENONnT, LZ and DARWIN, it extends well below the region where coherent neutrino-nucleus scattering will pose a serious source of background of a future generation of potentially much larger direct detection experiments.

## 5 Indirect Detection

WIMPs could also be detected by observing the radiation produced when they annihilate or decay [60, 61]. The flux of annihilation products is proportional to $(\rho_W/m_W)^2$, thus regions of highest interest are those expected to have an enhanced WIMP concentration. Potential signatures are high energy neutrinos from the Sun's core and from the Galactic Centre, gamma-rays from the Galactic Centre and from dwarf spheroidal galaxies, and positrons, antiprotons and antideuterons from the galactic halo. The predicted fluxes depend on the particle physics model delivering the WIMP candidate and on astrophysical input such as the dark matter halo profile, the presence of sub-structure and the galactic cosmic ray diffusion model, the latter being particularly relevant for the propagation of charged particles. In addition, the backgrounds to a potential dark matter signal must be well understood. These depend heavily on the production cross sections and on the cosmic ray propagation model.

Several existing observational anomalies have been interpreted as signatures for dark matter annihilation or decay in our galaxy, or in extragalactic dark matter halos [62, 63]. Among these are a rising positron fraction in the PAMELA [64], and AMS [65] data, a potential excess in the antiproton to proton ratio over the expected background in the latest AMS data, an excess of ∼1-3 GeV gamma rays from the region surrounding the Galactic Centre [66–69] in the Fermi-LAT data, and an unidentified 3.5 keV line observed in the stacked X-ray spectrum of galaxy clusters [70] and in the X-ray spectra of the Andromeda galaxy and the Perseus cluster [71]. The recent AMS data can however be explained by a new estimate of the antiproton background [72], and some groups do not find evidence for the claimed X-ray excess in galaxies and clusters [73]. Most recently, the Fermi collaboration does not see a gamma ray excess in dwarf spheroidal Milky Way satellites, which are some of the most dark matter dominated galaxies [74].

Even under the bold assumption that some of the above anomalies are due to dark matter, there is clearly no single candidate capable of explaining all the data, as dark matter particle masses from 7 keV, to ∼9-50 GeV to several TeV would be required. There is a strong demand for more data, which will come from the continued operation of Fermi and AMS, from the current generation of atmospheric Cherenkov detectors and the future CTA [75], as well as from existing and future neutrino experiments. In addition, WMAP and Planck data provide direct constraints on dark matter annihilation that takes place during the era of recombination, for WIMP annihilation could give rise to a sufficient number of energetic particles to impact the observed anisotropies in the cosmic microwave background [1, 76]. As experimental sensitivities are thus poised to drastically improve during this decade, we urgently need a deeper understanding of the astrophysical backgrounds and of the dark matter distribution on galactic scales.

Where no excesses are seen, or under the assumption that these are of astrophysical nature involving baryons and not dark matter, bounds on the WIMP annihilation cross section as a function of WIMP mass can be derived. A number of observations start probing the region around $\langle \sigma_{ann} v \rangle = 3 \times 10^{-26}$ cm$^2$s$^{-1}$, which is the value for a simple thermal relic. An example is shown in Figure 3 (from [74]), for the case of diffuse gamma radiation from WIMP pair annihilation: Fermi-LAT already probes cross sections below the thermal relic value for WIMP masses ≲100 GeV after 6 years of observation of 15 dwarf spheroidal satellites of the Milky Way [74]. By observing the Galactic Centre, the future CTA will be able to probe annihilation cross sections below the canonical value for WIMP masses in the region 100 GeV-10 TeV, depending on the assumed annihilation channel and dark matter halo profile [77].

It is perhaps most straightforward to compare direct dark matter detection results with those from searches for high-energy neutrinos from the Sun, as the WIMP-proton cross section plays a crucial role, initiating the capture process. WIMPs with orbits passing through the Sun can scatter from nuclei and lose kinetic energy. If their final velocity is smaller than the escape velocity, they will be gravitationally trapped and will settle to the Sun's core. Over the age of the solar system, a sufficiently large number of particles can accumulate and efficiently annihilate, whereby only neutrinos are able to escape and be observed in terrestrial detectors. The primary annihilation spectrum is once again model dependent and the range of possible models is usually considered by assuming 100% branching into channels with different characteristics: the so-called hard channel, where the WIMPs annihilate into $W^+W^-$ and the so-called soft channel, with annihilation into $b\bar{b}$. Typical neutrino energies are 1/3–1/2 of the WIMP mass, thus well above the solar neutrino background.

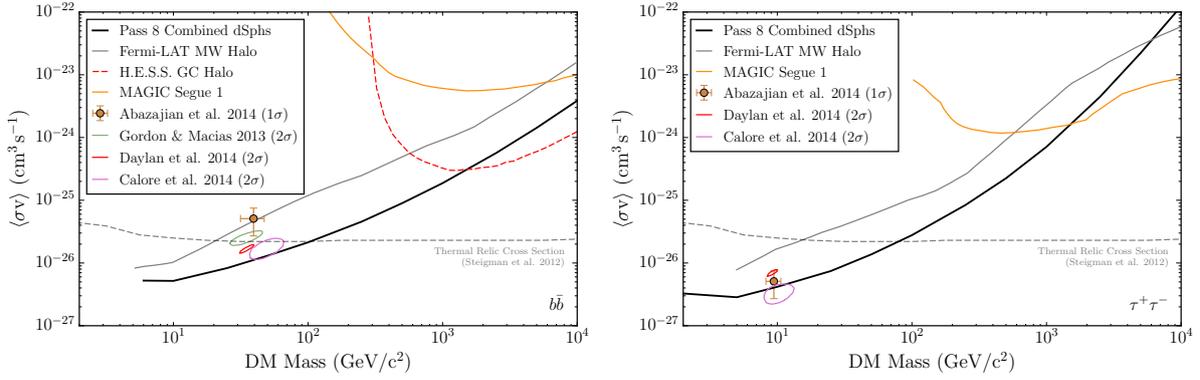

**Figure 3** Constraints on the WIMP annihilation cross section for the $b\bar{b}$ (left) and $\tau^+\tau^-$ (right) channels from Fermi-LAT [74, 78], HESS [79], and MAGIC [80]. The thermal relic cross section of $\langle \sigma v \rangle \approx 3 \times 10^{-26}$ cm$^3$s$^{-1}$, is shown as a dashed horizontal curve. Closed contours show the regions from several dark matter interpretations of the Galactic center excess [67–69, 81]. Figure from [74].

The strongest limits on high-energy neutrinos coming from the Sun are placed by IceCube [82, 83] and Super-Kamiokande [84, 85]. As an example, Figure 4 shows the upper limits on spin-dependent WIMP-proton cross section as a function of the WIMP mass (in the hard and soft annihilation channels) from IceCube [82], in comparison with results from direct detection experiments. In terms of the WIMP annihilation cross section, IceCube probes values around few$\times 10^{-20}$cm$^2$s$^{-1}$ - few$\times 10^{-24}$cm$^2$s$^{-1}$ for WIMP masses of 20 GeV-10 TeV, depending on the annihilation channel, as shown in Figure 4 [83].

## 6 Dark matter at the LHC

Another avenue to search for WIMPs ($\chi$) is to look for their production at the LHC: $pp \to \chi\bar{\chi}$. While the presence of dark matter particles is not directly observable in a detector at a collider, it can be inferred from their recoil on standard model particles. So far, there is no evidence for dark matter from LHC searches. In particular, there is no evidence for supersymmetry, or for any other new theoretical model motivated by solving the naturalness problem [6]. In more model-independent dark matter searches, the unknown interactions between dark matter and standard model particles are usually described by a set of effective operators. The expected signature is then missing transverse energy accompanied by a so-called mono-object (denoted by $X$: a photon, a single jet, a $Z$, etc) required to tag the event: $pp \to \chi\bar{\chi} + X$. Presently, the collider searches using mono-jets or mono-photons accompanied by missing transverse energy remained fruitless [88–90]. In addition, unlike the case for direct detection, the effective field theory (EFT) approach is not always valid, for instance when the energy scale probed by the effective operators is smaller than the energy of the partons taking part in the collision [91, 92].

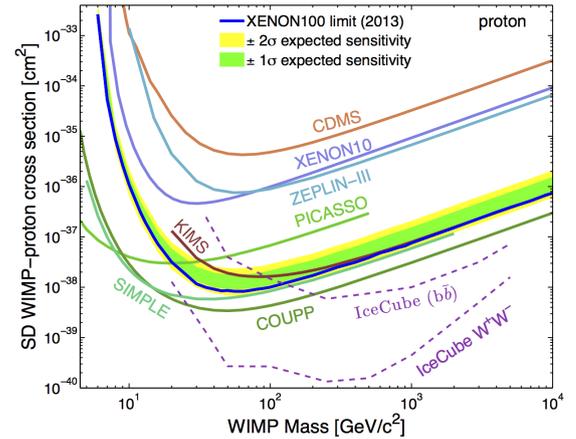

**Figure 4** IceCube [82] upper limits on the spin-dependent, WIMP-proton cross section, in the hard ($W^+W^-$, $\tau^+\tau^-$ for WIMP masses <80.4 GeV/c$^2$) and soft ($b\bar{b}$) annihilation channels. Also shown are results from direct detection experiments. Figure from [86].

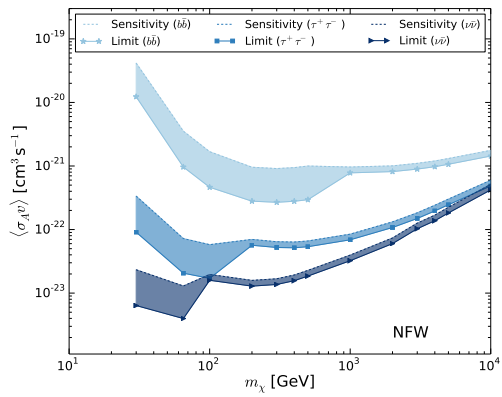

**Figure 5** IceCube sensitivities and upper limits on WIMP annihilation cross sections to $\tau^+\tau^-$, $b\bar{b}$ and to neutrinos, assuming a Navarro-Frenk-White (NFW) [87] dark matter halo profile. Figure from [83].

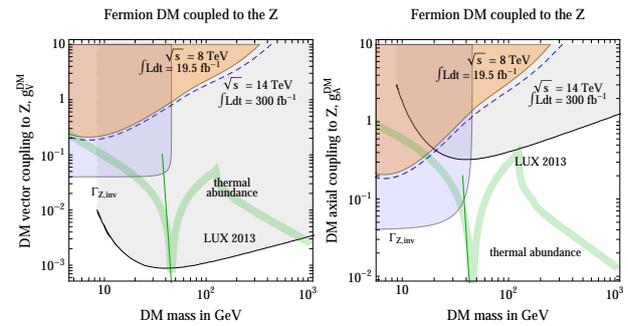

**Figure 6** Regions of the dark matter mass and vector ($g_V^{DM}$) and axial-vector ($g_A^{DM}$) couplings to the Z, along with constraints from the LHC, from direct detection (the LUX 2013 results) and from the Z-invisible width constraint. Also shown is the thermal relic abundance via Z-coupling annihilation. Figure from [92].

To allow a meaningful comparison with direct detection experiments, benchmark scenarios were proposed (see e.g., [91–93] and references therein). One approach is to classify possible mediators of the interaction between the dark matter and the standard model particles. For particles exchanged in the $s$−channel, the mediators must be electrically neutral, with spin 1 or 0, while for particles exchanged in the $t$−channel, the mediator can be a colour triplet. An example for an $s$−channel mediator is a new massive spin-one vector boson, $Z'$, from a broken U(1)' gauge symmetry. However, the mediator of interactions between the WIMP and quarks could also be the Z boson or the Higgs boson.

As an example which nicely illustrates the complementarity between collider and direct searches, Figure 6 (from [92]), shows the vector ($g_V^{DM}$) and axial-vector ($g_A^{DM}$) coupling versus mass regions for fermionic dark matter that couples to the Z. Shown are constraints from the LHC, from direct detection (exemplified by the LUX 2013 results) and from the Z-invisible decay width. Also shown is the predicted thermal relic abundance via Z-coupling annihilation, and the predictions for LHC14.

Another example for an alternative to the EFT interpretations is to characterise searches for dark matter via simplified models, that are constructed using four parameters only: the mass of the WIMP, the mass of the mediator and the couplings of the mediator to the WIMP, as well as to quarks [91, 93]. Assuming that the WIMP is a Dirac fermion and that the mediator couples to all quarks with the same strength, it is straightforward to compare collider and direct detection searches. For the exchange of a vector mediator, the LHC mono-jet searches have better sensitivity at WIMP masses below ∼5 GeV. For axial-vector mediators, the LHC has greater sensitivity than direct searches at WIMP masses below ∼200 GeV, as shown in Figure 7 (from [93]). The figure shows projected spin-dependent and spin-independent sensitivities in the WIMP-nucleon cross section versus WIMP mass for LHC14 (and various couplings) and a large liquid xenon detector (7 t LXe, exemplified by LZ), as well as a Si+Ge detector such as SuperCDMS. The region below which coherent neutrino-nucleus scattering will start do dominate the event rates in direct detection experiments is also shown.

## 7 Complementarity

The various dark matter search channels presented here are thus highly complementary to one another. While the current generation of collider searches can probe very low WIMP masses, and up to masses of ∼1 TeV, direct and indirect detection experiments can access the WIMP mass region up to 10 TeV and above. The complementarity is illustrated in Figure 8, for the concrete case of the CMSSM. Shown are the allowed regions in the $m_{1/2}$ versus $m_0$ parameter space, along with models that can be probed by the 14 TeV run at the LHC, by indirect detection for a diffuse $\gamma$-ray signal with CTA and by direct detection with XENON1T [53].

In terms of the WIMP couplings to standard model particles, colliders and high-energy neutrinos searches from the Sun are superior for axial-vector couplings, while direct searches show largest sensitivities for spin-independent scattering.

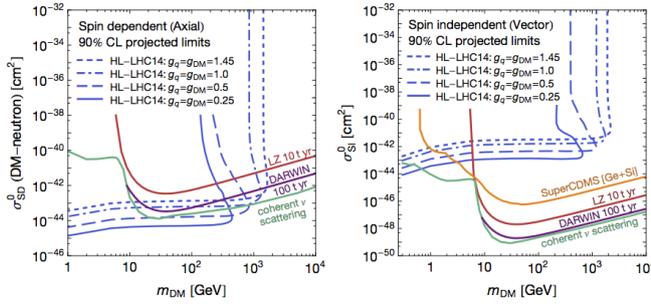

**Figure 7** Projected spin-dependent (left) and spin-independent (right) 90% CL limits for LHC14 in various coupling scenarios (blue lines) and for direct detection experiments (exemplified by large liquid xenon detectors, LZ (red line) and DARWIN (magenta line), and by a Si+Ge detector, SuperCDMS (orange line)). The region below which coherent neutrino-nucleus scattering will start do dominate the event rates in direct detection experiments is also shown (green curve). Figure from [93].

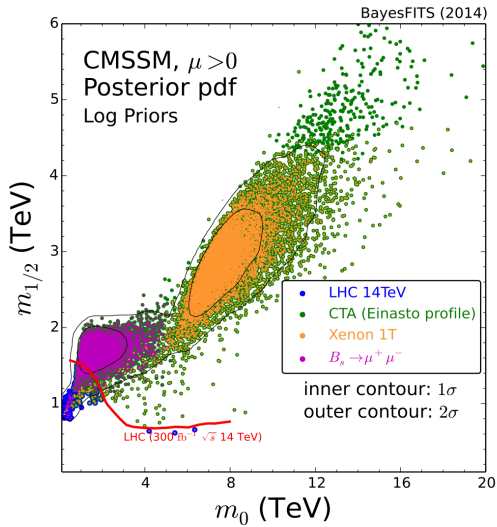

**Figure 8** Complementarity of LHC searches, direct detection (here XENON1T) and indirect searches (CTA) for a diffuse $\gamma$-ray signal in the case of the CMSSM. The allowed regions consider the Higgs mass of $\sim$125 GeV, lower limits from direct SUSY searches, and a relic WIMP abundance of $\Omega_{DM}h^2$ = 0.12. Figure from [53].

## 8 Outlook

Although invisible, dark matter is five times as abundant as normal matter in our Universe. Its identity, more than eighty years after its postulation by Fritz Zwicky in its modern form, remains a secret. Uncovering the nature of this dominant form of matter is thus one of the grand challenges of modern physics. Considering the immense progress in a large range of experiments operated at the Earth's surface, in space or deep underground, we might be faced with multiple and hopefully consistent discoveries within the next couple of years. In this case, the immediate next goal would be to reveal the detailed properties of dark matter particles, such as their mass, spin and couplings to ordinary matter, and to shed light on their phase-space distribution in the Milky Way's halo. This could herald the start of a new field, dark matter astronomy.

**Key words.** Keywords.